\documentstyle[12pt]{article}

\begin{document}

\begin{flushright}
IJS-TP-96/6\\
February 1996\\
\end{flushright}

\vskip 1cm

\begin{center}
{\Large \bf THE CONTROVERSY IN THE $\gamma\gamma\to\rho\rho$ 
PROCESS: POTENTIAL SCATTERING OR $qq\bar{q}\bar{q}$ RESONANCE ?}

\vspace{1.5cm}

{\large \bf Borut Bajc$^{a}$, Sa\v sa Prelov\v sek$^{a}$ 
and Mitja Rosina$^{a,b}$}

\vspace{.5cm}

{\it a) J. Stefan Institute, Jamova 39, P.O.B. 3000, 
1001 Ljubljana, Slovenia}

\vspace{.5cm}

{\it b) Faculty of Mathematics and Physics, University of 
Ljubljana, Jadranska 19, P.O.B .64, 1001 Ljubljana, Slovenia}

\vspace{1cm}

\end{center}

\centerline{\large \bf ABSTRACT}

\vspace{0.5cm}

The $\gamma\gamma\to\rho^0\rho^0\to 4 \pi$ reaction 
shows a broad peak at 1.5 GeV in the $(J^P,J_z)=(2^+,2)$ 
channel which has no counterpart in the $\rho^+\rho^-$ 
channel. This "resonance" is considered as a candidate 
for a $qq\bar q\bar q$ state in the "s-channel". We show, 
however, that it can also be explained by potential 
scattering of $\rho^0\rho^0$ via the $\sigma$- exchange 
in the "t-channel".

\vskip 1cm

\noindent
{\bf 1. Introduction}

\vskip 0.5cm

After many successes of quark models to describe a 
single meson or baryon, their predictive power in 
the two-hadron sector remains questionable. A fair 
description of the nucleon-nucleon interaction \cite{nnint} 
was accompanied with the prediction of a rich dibaryon 
spectrum which has never been observed; it is not yet explained
whether the predictions are wrong or the levels just too broad.
Therefore it is of great 
interest to see how different quark models perform in 
the two-meson sector, as compared to effective mesonic models.

Much experimental and theoretical interest has concentrated 
in the $\rho\rho$ system. Firstly, there is a remarkable 
experimental "puzzle". The $\gamma\gamma\to\rho\rho$ 
reaction is characterized by the following features 
\cite{argusone,argustwo}: a large and broad peak 
near the nominal threshold in the $\rho^0\rho^0 ~~~~ (2^+,2)$ 
channel and a much smaller cross section ("noise") in other 
$\rho^0\rho^0 $ channels as well as in the $\rho^+\rho^-$ channel. 

Secondly, the $\rho\rho$ system consists of two unstable 
(shortlived) hadrons. Therefore, the shape of the 
"resonance" is unusual and it challenges an appropriate 
quantum mechanical treatment. Also, such shape might
be selective regarding competing models. More experimental 
data would be very welcome in order to sharpen 
the details of this shape.

A good model should explain all these features simultaneously. 
Since so far the quark models \cite{achasov,liliu} covered 
more features simultaneously than the effective mesonic models 
\cite{moussallam,hatzis}, the opinion prevails that explicit 
quark degrees of freedom (a $qq\bar{q}\bar{q}$ molecule) are 
essential for this process. In the present paper we challenge 
this view. We show that the considered process can also be 
explained by potential scattering.

Of course, also in potential scattering, the intermediate 
system consists of two quarks and antiquarks, each $q\bar{q}$ 
pair being correlated like in a free $\rho$ meson. This is 
just one important configuration, which also extends 
into the asymptotic region. Other intermediate 
$qq\bar{q}\bar{q}$ configurations (such as those suggested 
in \cite{achasov,liliu}) may or may not be important. 
They can easily lead to a resonance peak but this does 
not prove that they really dominate (that they are at 
the right energy and strongly coupled to the asymptotic 
wavefunction) unless no other mechanism exists. 

The peak near the nominal threshold and its strong 
spin dependence are reminiscent of the p-n scattering 
at low energy. There the S=0 channel has an exceptionally 
high cross section while in the S=1 channel it is much lower. 
The reason is that in the S=0 state the potential well 
contains almost exactly 1/4 wavelength of the relative 
wavefunction while in the S=1 state there is a little 
 more than $\lambda/4$ in the well because the potential 
is slightly stronger. 
We hypothesize that a similar trick of nature is 
played also in the case of the $\rho-\rho$ scattering. 

We assume that both photons are converted in $\rho$ 
via vector dominance, and both $\rho$ then interact 
by a scalar isoscalar potential which is just strong 
enough to (almost) bind them for one spin orientation 
and fails to do so for the other spin orientations. 
In addition, an isoscalar potential has the desired 
feature that it cannot exchange charges and does not 
lead to the $\rho^+\rho^-$ final state.

\vskip 1cm

\noindent

{\bf 2. A nonrelativistic potential model}

\vskip 0.5cm

In order to illustrate this idea, we choose a Yukawa-type
$\sigma$-exchange potential $V(r)=V_0 \exp(-m_\sigma r)/
(-m_\sigma r)$.  The full process
is shown in Fig. 1 and its  Lorentz invariant 
amplitude can be factorized in the following way:

\begin{equation}
M=M_{\gamma\gamma\rho\rho}~
P_{\rho 1}~P_{\rho 2}~f_{\rho\pi\pi}^2\;.
\label{eqone}
\end{equation}

We treat the $\gamma\to\rho$ conversion and $\rho\to\pi\pi$ 
decay perturbatively, but the $\rho\rho\to\rho\rho$ scattering 
nonperturbatively by solving the nonrelativistic Schr\"odinger 
equation. In the calculation of the matrix element for the 
process $\gamma_1 ~ \gamma_2 \to \rho_1 ~\rho_2$ we account for 
inelastic scattering of $\rho$-mesons: 

\begin{equation}
M_{\gamma\gamma\rho\rho}=\int 
\psi _{\gamma\gamma}(\vec r_{\gamma 1},\vec r_{\gamma 2})~
\hat V_{\gamma\rho}~
\psi _{\rho\rho}(\vec r_{\rho 1},\vec r_{\rho 2})~ 
d\vec r_{\gamma 1}d\vec r_{\gamma 2}
d\vec r_{\rho 1}d\vec r_{\rho 1}
\label{Mgammarho}
\end{equation}

\noindent
where $\hat V_{\gamma\rho}$ is the contact $\gamma\rho$ potential 
$\hat V_{\gamma\rho}=(f_{\gamma\rho}^2/ M_{\rho}^3)~
\delta^3 (\vec r_{\gamma 1}-\vec r_{\rho 1})
\delta^3 (\vec r_{\gamma 2}-\vec r_{\rho 2})$,
$\psi_{\gamma\gamma}$ is the function of incident photons (plain 
wave, normalized in a box of volume $V$) 
and $\psi_{\rho\rho}$ is the wave function of the scattered $\rho$ 
mesons, calculated by solving the Schr\"odinger equation. The amplitude 
for $\gamma\to\rho$ conversion $f_{\gamma\rho}$ is determined from 
the experimental value for the partial width of the process 
$\rho\to e^+e^-$ \cite{pdg} and is equal to 
$f_{\gamma\rho}^2/ m_{\rho}^4=(2.01*137)^{-1}$.
In eq. (\ref{Mgammarho}) the photon selects the correct momentum 
component (Fourier component) from the calculated scattering wave 
function of the two mesons. 

We account for the shortlived nature of $\rho$ mesons  ($\Gamma=152 $ MeV) 
while writing the propagators of 
the scattered mesons $P_{\rho 1}$ and $P_{\rho 2}$:

\begin{equation}
P_{\rho i}={1\over W_{\rho i}^2-{\vec p}^2-(m_{\rho}-
{1\over 2}i\Gamma )^2}~~.
\end{equation}

We reduce the Lorentz invariant four-body phase space $d\Phi _4$ 
\cite{pdg} to two-body phase space of
the two $\rho$ mesons by integrating over all pion momenta
and assuming that the amplitude is not sensitive to them.
Neglectig the contributions of order $\Gamma^2/m_{\rho}^2$ we get 
the cross-section

\begin{equation}
\label{eqtwo}
\sigma (W)=\int {(2\pi)^4 \vert M\vert ^2 d\Phi _4\over 2 W^2}
=\int dm_{\rho 1}^2~\int dm_{\rho 2}^2
~{1\over (2\pi)^3}~\vert M_{\gamma\gamma\rho\rho}\vert^2
~{\vert {\vec p}\vert \over W^3}~
\end{equation}

$$\times{m_{\rho 1} \Gamma \over
(m_{\rho 1}^2-m_{\rho}^2)^2+m_{\rho}^2\Gamma ^2}~
{m_{\rho 2} \Gamma \over 
(m_{\rho 2}^2-m_{\rho}^2)^2+m_{\rho}^2\Gamma ^2}~~,$$

\noindent
where $m_{\rho 1}$ and $m_{\rho 2}$ are the effective masses of 
the scattered $\rho$-mesons $m_{\rho i}^2=W_{\rho i}^2-\vec p^2$. 
This formula (used frequently before) has an intuitive 
interpretation that the two "final" $\rho$ mesons have 
arbitrary masses with a Breit-Wigner probability 
distribution around the nominal mass $m_\rho=770 $ MeV.

Since the scattering phase shift is very sensitive to 
potential parameters as well as on $\rho$ masses we do 
calculate it separately for each combination of the two 
$\rho$ masses and add up the contributions in the double 
integral numerically, with the appropriate Breit-Wigner weights.

The calculated cross sections $\sigma(W)$ are very sensitive 
to potential strenght $V_0$ and weakly sensitive to potential 
range $1/m_{\sigma}$ (in this nonrelativistic model, 
$m_{\sigma}=550 $ MeV is taken). Results for different 
$V_0$ are presented in Fig. 2. 

For a  narrow range of the model parameter $V_0$, the
results qualitatively agree with the experimental data, 
as long as we are considering the height of the peak 
in the $(2^+,2)$ channnel. The position of the peak comes, 
however, too high above 
the nominal threshold. Similar calculation using relativistic 
Klein-Gordon equation for the $\rho$-meson scattering on fixed 
potential shows, that the tendency of  relativistic effect is to 
pull the peak down to lower energies. It will be, however, shown 
in Sect. 3 that a more severe calculation including  
relativistic effects and gauge invariant coupling can cure this
deficiency of the toy model.

In order to get a very low cross section in the $(2^+,0)$ 
channnel one has to assume a strong spin dependence of 
the $\sigma$-exchange potential. Such strong spin dependence is 
arbitrary in the present nonrelativistic model, but 
it will in fact appear in the relativistic model in 
Sect.3 as a consequence of momentum dependence of the gauge 
invariant coupling.

\vskip 1cm

\noindent
{\bf 3. A relativistic model with gauge invariant 
$\rho\rho\sigma$ coupling}

\vskip 0.5cm
The Lagrangian, from which we obtain the potential, 
respects vector dominance and is gauge invariant:

\begin{eqnarray}
\label{lagrangian}
{\cal L}=&-&{1\over 4}F_{\mu\nu}(B)F^{\mu\nu}(B)
-{1\over 4}F_{\mu\nu}(\rho)F^{\mu\nu}(\rho)
+{1\over 2}m_\rho^2\rho_\mu^2
+{1\over 2}(\partial_\mu\sigma)^2-{1\over 2}m_\sigma^2\sigma^2\nonumber\\
&+&{g_\sigma\over 2}\sigma[F_{\mu\nu}(\rho)F^{\mu\nu}(\rho)-
\alpha {e \over g} F_{\mu\nu}(\rho)F^{\mu\nu}(B)-
\beta ({e \over g})^2 F_{\mu\nu}(B)F^{\mu\nu}(B)]\nonumber\\
&-&{1\over 2}m_\rho^2[\rho_\mu^2-(\rho_\mu-
{e\over g}B_\mu)^2]\;,
\end{eqnarray}

\noindent
where $F_{\mu\nu}(\Phi)=\partial_\mu\Phi_\nu-\partial_\nu\Phi_\mu$. 
After diagonalyzing the quadratic term in the Lagrangian, 
we may want to have a vanishing $\sigma\gamma\gamma$ vertex, 
as approximately required by the low-energy 
$\gamma\gamma\to\pi\pi$ scattering and confirmed in some 
phenomenological models as the NJL model with a 
sharp cutoff \cite{ournjl}. This fixes $\beta=1-\alpha$. 
Such an assumption is however not needed in our 
calculation, since our result in the leading order of 
$e/g$ does not depend on the parameter $\beta$. 
 
The unitary amplitude is calculated with the Bethe-Salpeter 
equation with the kernel given by Eq. (\ref{lagrangian}). 
The equation is simplified by the ladder approximation 
(only $\sigma$ exchange in the $t$ and $u$ channels) and
by taking into account only the leading contributions 
in powers of $e/g$. The Blankenbecler-Sugar 
reduction \cite{blankenbecler} is used (with no retardation, 
i.e. with equal off-shellness of the final and 
intermediate $\rho$) and the partial wave 
expansion \cite{jacob}, \cite{gross} is performed: 

\begin{eqnarray}
\label{bsequation}
M^J_{{\lambda'}_1{\lambda'}_2\lambda_1\lambda_2}(q',q) & = & 
V^J_{{\lambda'}_1{\lambda'}_2\lambda_1\lambda_2}(q',q) + 
\int {dk \over k^2-q_m^2-i\epsilon} \times \nonumber \\
& & V^J_{{\lambda'}_1{\lambda'}_2{\lambda''}_1{\lambda''}_2}
(q',k) f_{{\lambda''}_1{\lambda''}_2}(k)
M^J_{{\lambda''}_1{\lambda''}_2\lambda_1\lambda_2}(k,q)\;,
\end{eqnarray}

\noindent
with

\begin{eqnarray}
\label{deff}
f_{{\lambda''}_1{\lambda''}_2} & = & 
{k^2 \over (4\pi)^3 \sqrt{k^2+m^2}} 
a_{{\lambda''}_1}({\tilde m}^2(k)) 
a_{{\lambda''}_2}({\tilde m}^2(k)) \;,
\end{eqnarray}

\noindent
where ${\tilde m}^2(x)\equiv W^2/4-x^2$ and 
the factor

\begin{equation}
\label{deffaktora}
a_\lambda(q^2) = \left\{\begin{array}{lll}
- & 1 & \qquad q^2<0 \quad \hbox{and} \quad \lambda=0 \cr
  & 0 & \qquad q^2=0 \quad \hbox{and} \quad \lambda=0 \cr
+ & 1 & \qquad \hbox{otherwise} \cr 
\end{array} \right.
\end{equation}

\noindent
takes into account that the longitudinal polarization 
vector $\epsilon_0^\alpha(q)$ becomes timelike (lightlike) 
for $q^2<0$ ($=0$). 

In eq. (\ref{bsequation}) the initial particles with 
polarizations $\lambda_1$ and $\lambda_2$ have 
the linear momentum $q$ and are on-shell 
($p_1^2=p_2^2=W^2/4-q^2=0$). The final (intermediate) 
$\rho$ mesons with polarizations $\lambda_1'$, $\lambda_2'$ 
($\lambda_1''$, $\lambda_2''$) can be equally off-shell 
with $p_3^2=p_4^2=W^2/4-{q'}^2$ ($W^2/4-k^2$). If the 
total energy $W$ is larger than the nominal threshold 
$2m$, then $q_m^2\equiv W^2/4-m^2>0$ and eq. 
(\ref{bsequation}) has complex solutions as required by 
unitarity.

Taking special linear combinations of amplitudes 
with definite polarizations one gets 
amplitudes with definite parity. Each channel 
gets decoupled from the others and can be 
represented by a system of one-dimensional integral 
equations. The system for the $(2^+,2)$ channel 
is explicitly given in Appendix A. 
These equations were then solved 
with the matrix inversion method \cite{brown}, 
taking $24$ points for the magnitude of the 
relative three-momentum from $0$ to the cutoff $\Lambda$. 
It was also checked, that the results change 
slightly, if one takes only $16$ grid points (the 
differences get however larger with energy, that 
is why we limited our calculation up to $W=1.8$ GeV). 

The cross section was then calculated by weighing the 
amplitude squared with two Breit-Wigner factors (one for each 
final $\rho$). The approximation of equal off-shellness 
of the two $\rho$, with which the Bethe-Salpeter equation 
(\ref{bsequation}) was derived and solved, is equivalent to 
the approximation that the amplitude depends 
only on the sum of the "masses" of the final $\rho$. 
For the total cross-section we get 

\begin{eqnarray}
\label{crsec}
\sigma_{TOT}&=&
\int_0^{{\tilde m}(2m_\pi)}dq' \rho(q')
{1\over 4}\sum_{J,\lambda_i^{(')}}(2J+1)
|M_{{\lambda'}_1{\lambda'}_2\lambda_1\lambda_2}^
{(J)}(q',q)|^2\;,
\end{eqnarray}

\noindent
where a first integration is already hidden in 

\begin{eqnarray}
\label{defrho}
\rho(q')&=&K({q'\over W^2})^2\int_0^{{\tilde m}(q')-2 m_\pi}dp'
[{\tilde m}^2(p')+{\tilde m}^2(q')-{\tilde m}^2(0)]
{{\tilde m}(p')\over{\tilde m}(q')}
\times\nonumber\\
& &F_{BW}(({\tilde m}(q')+p')^2,m_\rho^2)
F_{BW}(({\tilde m}(q')-p')^2,m_\rho^2)\;,
\end{eqnarray}

\noindent
where $K=[(e/g)(1-\alpha/2)]^4/(8\pi^5)$. 
The relativistic Breit-Wigner factor is defined as 

\begin{equation}
\label{fbw}
F_{BW}(p^2,m_\rho^2)={m_\rho\Gamma_\rho(p^2)\over 
(p^2-m_\rho^2)^2+m_\rho^2\Gamma_\rho^2(p^2)}
\end{equation}

\noindent
with the same $p^2$-dependent $\rho$-meson decay width 
\cite{jackson} as in the experimental analysis \cite{argusone}:

\begin{equation}
\label{sirinaro}
\Gamma_\rho(p^2)=\Gamma_\rho(m_\rho^2)
({p^2-4m_\pi^2\over m_\rho^2-4m_\pi^2})^{3\over 2} 
{2(m_\rho^2-4m_\pi^2)\over (p^2-4m_\pi^2)+(m_\rho^2-4m_\pi^2)}\;.
\end{equation}

\noindent
with $m_\rho=770$ MeV, $\Gamma_\rho(m_\rho^2)=152$ MeV 
and $m_\pi=140$ MeV. 

The cross sections in separate channels are obviously 
only part of the total cross section (\ref{crsec}). 
The explicit expression for the $(2^+,2)$ channel is 
given in Appendix A.

We take for the vector dominance factor $e/g=$ 
from $g^2/(4\pi)=2.3$ \cite{poppe}. The remaining parameter 
$\alpha$ were fitted to the experimental data in 
order to get the correct height of the cross section 
(the position of the peak and the relative contributions 
of different channels are however not dependent on 
$\alpha$). We have checked that the fitted value for 
$\alpha$ gives a small enough ($10^{-4}$) 
branching ratio for $\rho^0\to\gamma\pi^0\pi^0$ 
decay, which is not observed experimentally. 
The mass in the two-body propagator after the 
Blankenbecler-Sugar 
reduction would be $m=m_\rho=0.77$ GeV for non-decaying 
$\rho$, but can be slightly modified in the more 
realistic case of broad $\rho$ mesons. We used 
the value $m=0.692$ GeV. Since we are interested in the 
energy range $1.4$ GeV$<W<1.8$ GeV and since $m<W/2$ the 
obtained amplitudes are unitary. We checked that 
the cross sections behave similarly 
for $W>2m$, if a different value for $m$ is chosen. 
It came also out, that the results are relatively 
insensitive to the choice of the $\sigma$ mass, 
so that we will show only the cross sections for 
$m_\sigma=0.5$ GeV. 

The results for the interesting channels are shown 
in Fig. 3, for different choices of the free model 
paramaters $\Lambda$, $g_\sigma$ and $\alpha$ together 
with the experimental values. We checked also that 
there is a combination of the three free parameters 
in a reasonable range, which reproduces quite well 
the experimental results: if we fix one of them, 
the other two are rather precisely constrained by the data.

\vskip 1cm

\noindent
{\bf 4. Conclusion}

\vskip 0.5cm

We have shown that there exist plausible sets of parameters 
for potential scattering (for the $\rho\rho\sigma$ 
coupling) such that the peak in the 
$\rho^0\rho^0$ $(2^+,2)$ channel at $1.4-1.8$ GeV 
is reproduced while in other channels the cross 
section remains small. The parameters are plausible 
if compared to the $\sigma$-exchange contribution 
of the N-N interaction (for example in Bonn 
potential \cite{holinde}). The agreement does not 
prove that the mechanism is just potential scattering, 
but it presents potential scattering as a candidate 
until proven otherwise. Of course, we neither prove 
that resonance scattering through a $qq\bar{q}\bar{q}$ 
molecule is wrong, we only show that it is not conclusive 
since other mechanisms exist. A finer distinction will be needed. 

The succes of the potential scattering relies on the choice of
model parameters which "hold" about a quarter of the wavelength
of the scattered $\rho$ in the "potential well", a situation
similar to the singlet $n-p$ scattering 
where there is also almost exactly a quarter wavelength in the well 
yielding a high peak at low energy.
The nonrelativistic toy model illustrates this point qualitatively;
the position of the peak comes, however, too high above 
the nominal threshold. 
Relativistic effects and gauge invariant momentum dependent
coupling  cure this deficiency; they also give a strong spin dependence of 
the $\sigma$-exchange potential needed to keep the cross-section low in
other channels. 

\vskip 0.5cm
This work was supported by the Ministry of Science and 
Technology of Slovenia.

\vskip 1cm

\noindent
{\bf APPENDIX A}

\vskip 0.5cm

In this Appendix we will write down some useful 
formulas for the calculation of the amplitudes 
and cross sections. We will concentrate only on 
one specific channel, namely, the most 
important $(J^P,J_z)=(2^+,2)$ channel.
Similar equations can be (and were) defined also 
for the other possible channels.

First one has to project the potential to states 
with definite total angular momentum. This was 
done integrating the potentials for $\rho\rho$ 
scattering obtained from the Lagrangian (\ref{lagrangian}) 
over the scattering angle \cite{jacob},\cite{gross}:

\begin{equation}
\label{vj}
V^{(J)}_{{\lambda'}_1{\lambda'}_2\lambda_1\lambda_2}(k',k)=
2\pi\int_{-1}^1d(cos\theta)V_{{\lambda'}_1{\lambda'}_2
\lambda_1\lambda_2}(\vec{k'},\vec{k})\;.
\end{equation}

Using these quantities one could solve for each $J$ 
the Bethe-Salpeter equation (\ref{bsequation}) for 
the $4\times 9=36$ unknown functions (of $q'$) 
$M^{(J)}_{\lambda'_1\lambda'_2\lambda_1\lambda_2}$. 
However, due to parity and Bose symmetry, not 
all these functions are independent. In fact, 
they are related by

\begin{eqnarray}
\label{relx}
X_{{\lambda'}_1{\lambda'}_2\lambda_1\lambda_2}^{(J)} (q',q)& = & 
X_{-{\lambda'}_1-{\lambda'}_2-\lambda_1-\lambda_2}^{(J)} (q',q)\;,\\
X_{{\lambda'}_1{\lambda'}_2\lambda_1\lambda_2}^{(J)} (q',q)& = & 
X_{{\lambda'}_2{\lambda'}_1\lambda_2\lambda_1}^{(J)} (q',q)\;,
\end{eqnarray}

\noindent
($X=M$ or $V$) 
so that the matrices $M(q',q)$, $M(k,q)$ in $V(q',q)$ 
in (\ref{bsequation}) have only $3$ independent 
elements for $J=0$ and $12$ for $J=2$, while $V(q',k)$ 
has $5$ independent elements for $J=0$ and $25$ for 
$J=2$ (the difference comes from the fact that the 
matrix elements of $V(q',k)$ can have also longitudinally 
polarized particles in the ket, while only transversal 
polarizations are allowed for the other, since 
$W^2/4-q^2=0$).

All the elements $X_{{\lambda'}_1{\lambda'}_2
\lambda_1\lambda_2}^{(J)} (q',q)$ do not have 
good parity. Using the equations (\ref{relx}) and 
taking special linear combinations we can decouple 
the original Bethe-Salpeter equation (\ref{bsequation}), 
obtaining lower dimensional matrix integral 
equations, one for each $(J^P,J_z)$. 

For example, we get for $(J^P,J_z)=(2^+,2)$ 
the following integral equation:

\begin{eqnarray}
\label{twoplustwo}
\pmatrix{     {M}^{(2)}_{+++-}  \cr
          \bar{M}^{(2)}_{+-+-}  \cr
          \bar{M}^{(2)}_{+0+-}  \cr
              {M}^{(2)}_{00+-}  \cr }(q',q) &=&
\pmatrix{     {V}^{(2)}_{+++-}  \cr
          \bar{V}^{(2)}_{+-+-}  \cr
          \bar{V}^{(2)}_{+0+-}  \cr
              {V}^{(2)}_{00+-}  \cr }(q',q) + 
\int{dk\over k^2-q_m^2-i\epsilon}\times\\
&&\pmatrix{ \bar{V}^{(2)}_{++++} & 
                {V}^{(2)}_{+++-} &
            \bar{V}^{(2)}_{+++0} & 
                {V}^{(2)}_{++00} \cr
                {V}^{(2)}_{+-++} & 
            \bar{V}^{(2)}_{+-+-} & 
            \bar{V}^{(2)}_{+-+0} & 
                {V}^{(2)}_{+-00} \cr
            \bar{V}^{(2)}_{+0++}       & 
            \bar{V}^{(2)}_{+0+-}       & 
            \bar{\bar{V}}^{(2)}_{+0+0} & 
                {V}^{(2)}_{+000}       \cr
            V^{(2)}_{00++} & 
            V^{(2)}_{00+-} & 
            V^{(2)}_{00+0} & 
            V^{(2)}_{0000} \cr }(q',k)
\times\nonumber\\
&&\pmatrix{2 f^+ & 0     & 0     & 0   \cr
           0     & 2 f^+ & 0     & 0   \cr
           0     & 0     & 4 f^- & 0   \cr
           0     & 0     & 0     & f^+ \cr}(k)
\pmatrix{     {M}^{(2)}_{+++-}  \cr
          \bar{M}^{(2)}_{+-+-}  \cr
          \bar{M}^{(2)}_{+0+-}  \cr
              {M}^{(2)}_{00+-}  \cr }(k,q)\;,\nonumber
\end{eqnarray}

\noindent
where we defined two averages 

\begin{eqnarray}
\label{averageone}
\bar{X}^{(J)}_{\lambda'_1\lambda'_2\lambda_1\lambda_2}&=&
{1\over 2}(X^{(J)}_{\lambda'_1\lambda'_2\lambda_1\lambda_2}+
X^{(J)}_{-\lambda'_1-\lambda'_2\lambda_1\lambda_2})\;,\\
\label{averagetwo}
\bar{\bar{X}}^{(J)}_{\lambda'_1\lambda'_2\lambda_1\lambda_2}&=&
{1\over 2}(\bar{X}^{(J)}_{\lambda'_1\lambda'_2\lambda_1\lambda_2}+
\bar{X}^{(J)}_{\lambda'_2\lambda'_1\lambda_1\lambda_2})\;,
\end{eqnarray}

\noindent
and where 

\begin{eqnarray}
\label{deffpm}
f^\pm(k)&=&{k^2a^\pm(W^2/4-k^2)\over (4\pi)^3 \sqrt{k^2+m^2}}
\end{eqnarray}

\noindent
and $a^+(x)=+1$, $a^-(x)=+1,0,-1$ for $x=+1,0,-1$ 
respectively.

The cross section in this same channel 
is then

\begin{eqnarray}
\label{pkdodatek}
{1\over\rho(q')}{d\sigma^{(2^+,2)}\over dq'}&=&
5(|{M}^{(2)}_{+++-}(q',q)|^2+
 |\bar{M}^{(2)}_{+-+-}(q',q)|^2+\nonumber\\
&& 2|\bar{M}^{(2)}_{+0+-}(q',q)|^2+
 {1\over 2}|{M}^{(2)}_{00+-}(q',q)|^2)\;.
\end{eqnarray}

\newpage

\centerline{FIGURES}

\vskip 0.5cm

\noindent
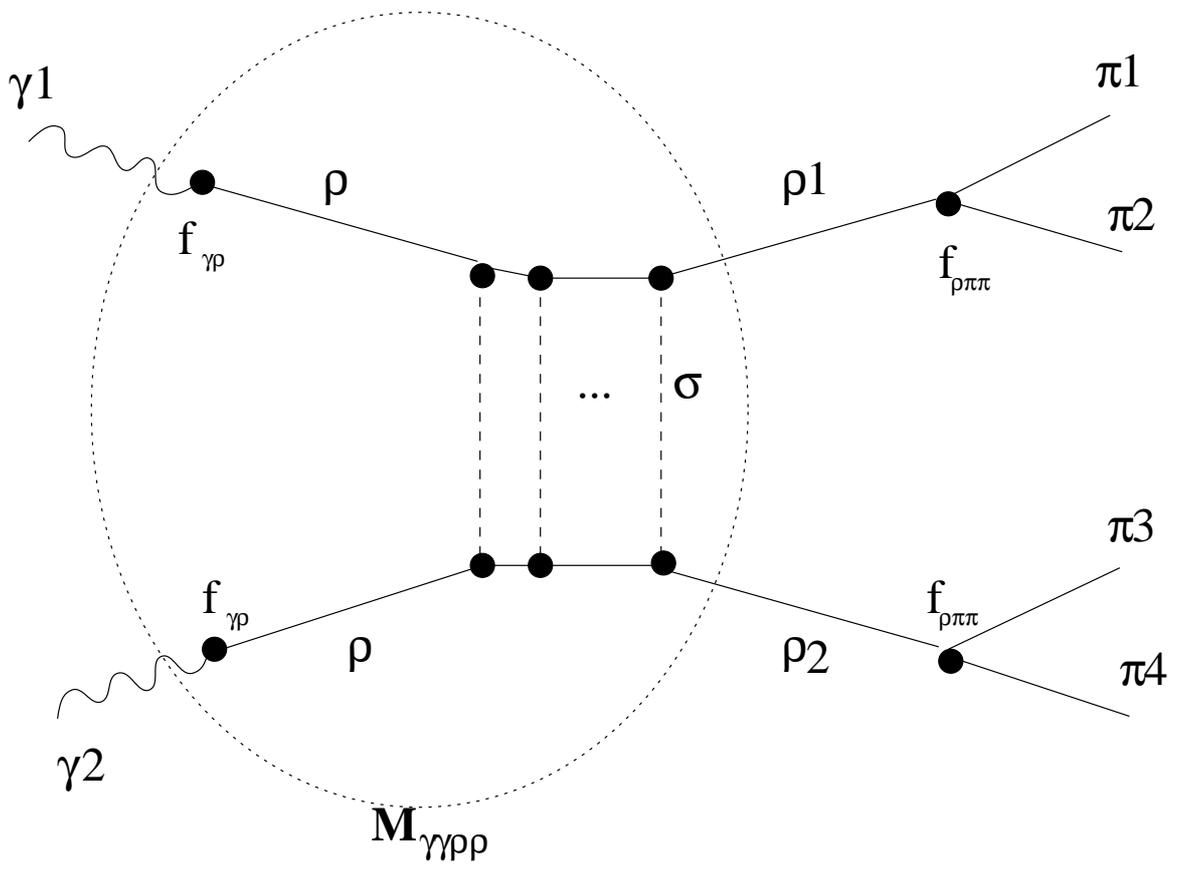
Figure 1: Feynman diagram for the process $\gamma\gamma\to\rho^0\rho^0\to
\pi\pi\pi\pi$.

\vskip 0.5cm

\noindent
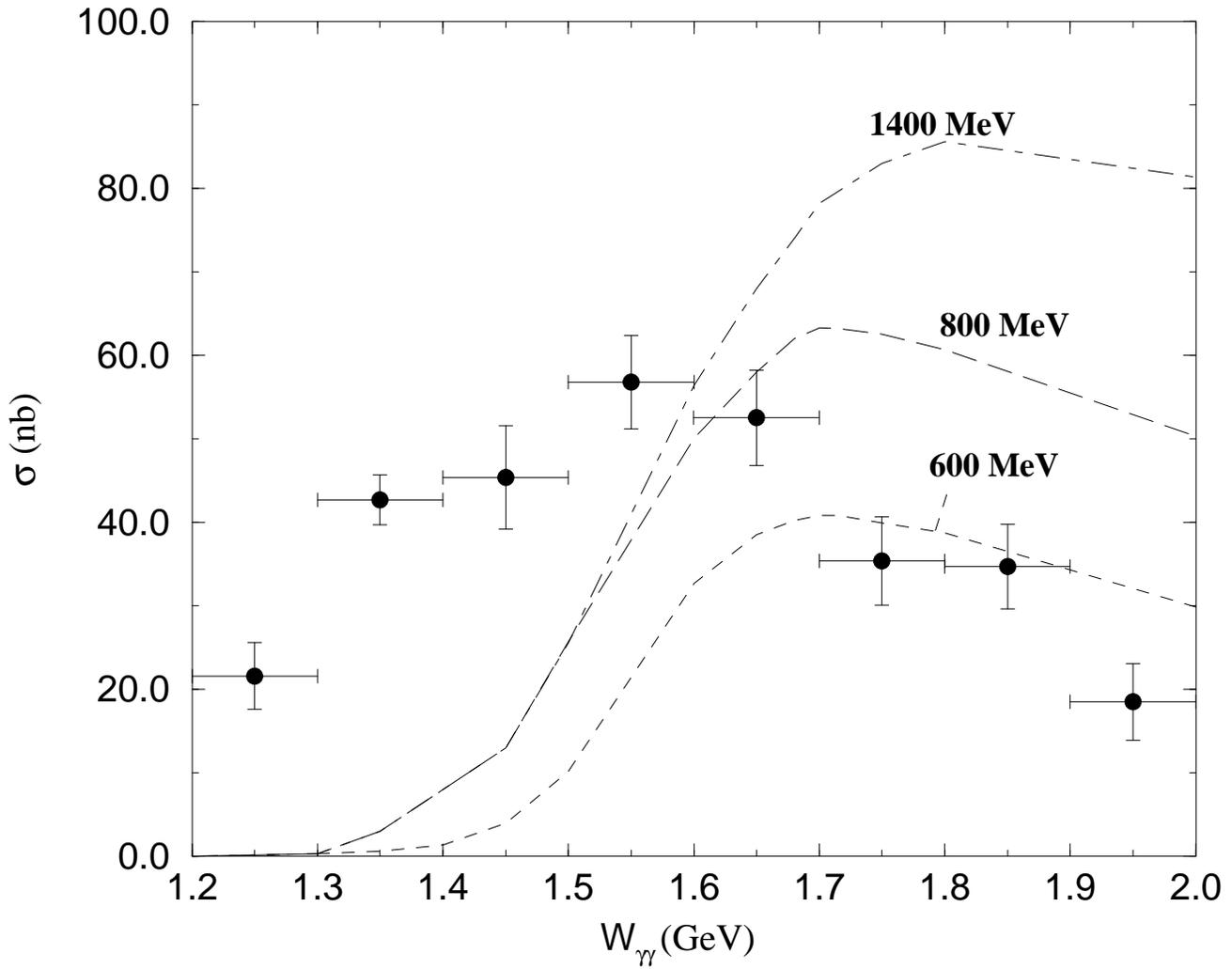
Figure 2: The  experimental cross sections for  the $(2^+,2)$ channel
\cite{argusone} (dots) 
and calculated cross sections for three different potential strengths 
$V_0$: $V_0=1400$ MeV gives the highest cross section, while 
$V_0=800$ MeV gives the cross section with the same height as  
the experimental one.

\vskip 0.5cm

\noindent
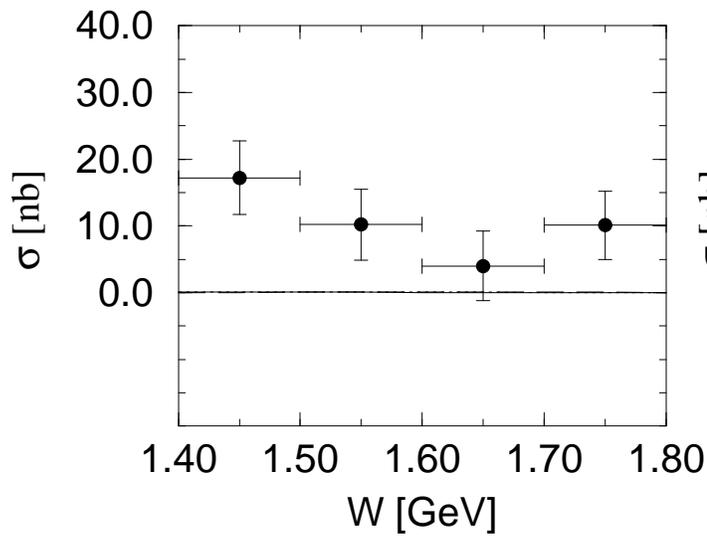
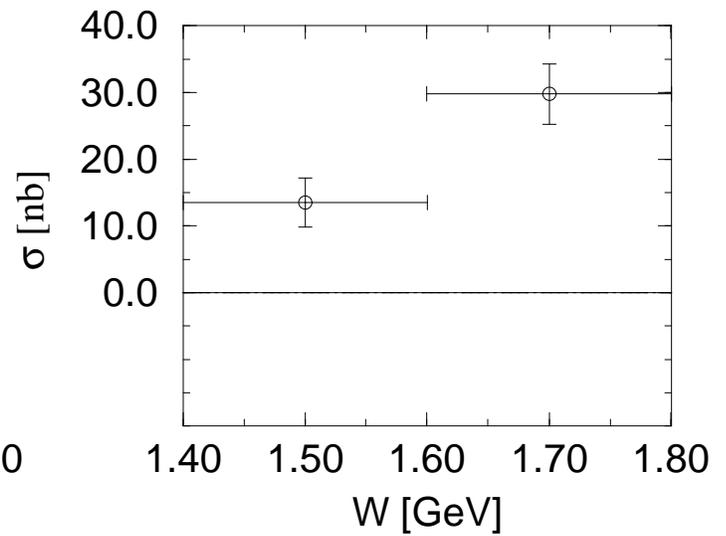
Figure 3: Cross sections for various channels in the relativistic 
potential model with scalar exchange in comparison with 
experimental data.

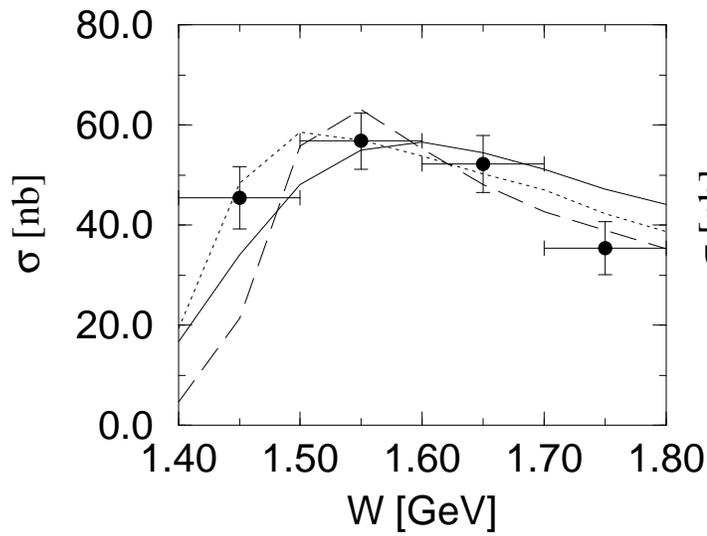
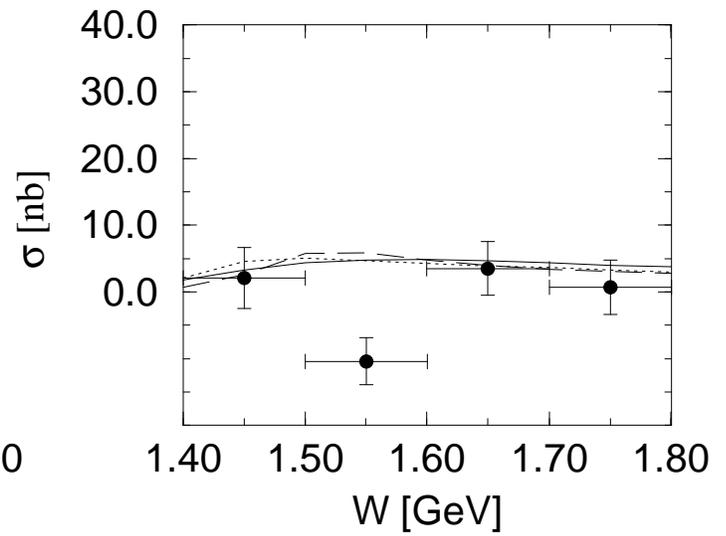
\end{document}